\documentclass{article}
\usepackage{algorithm}
\usepackage[indent=15pt, parfill]{parskip}
\usepackage{indentfirst}
\title{Selective Algorithm Processing of Subset Sum Distributions}
\author{Nick Dawes, Ottawa, Canada: nwdawes314@gmail.com}
\date{September 16, 2024}

\begin{document}

\maketitle

\section {Abstract}

One form of the subset sum problem is to decide if there is a subset of an input multiset of $n$ positive integers that exactly sum to $T$, a target integer. This and other previously reported systems solve this form of this problem. This system has experimentally validated computational O(max($T$, $n^2$)).
The computing efficiency of systems which solve this problem and which compute individual subset sums is $e=min(T/z, 1)$, where $z$ is the number of subset sums computed.
$e$ is related to these system's computational complexity. This system optimises its efficiency and so reduces its computational complexity. It maps the $T$ sums into $kn$ equal sized bins, where $k$ is a small constant. This mapping lets it select and apply its most efficient algorithm for each bin and an input value to the sums in that bin for that input value.
Its selectable algorithms include additive, subtractive and repeated value dynamic programming. 
Cases which would otherwise be processed inefficiently (eg: all even values) are handled by modular arithmetic and by dynamically partioning the input values. 
The system's efficiency corresponds to computational complexity O(max($T$, $n^2$)). Its computational complexity is experimentally validated but remains unproven. The system has space complexity O(max($T$, $n$)).

\textbf {Acknowledgments:} I thank Michel Ranger for his comments on this paper.

\section {Introduction}

One form of the subset sum problem is to decide if there is a subset of a multiset of $n$ positive integers $v_i: i=1..n$ that exactly sum to $T$, a target integer [KA72].  This form of this problem is NP-complete.
There are two main classes of algorithms which solve it: exponential time and pseudopolynomial time. Horowitz and Sahni [HS74] first described a divide and conquer algorithm with exponential time, O($2^{n/2}{n/2}$). Schroeppel and Shamir[SS81] improved this to O($2^{n/2}{n/4}$) with much smaller space complexity.  Bellman's classic dynamic programming [BE57] can be applied in pseudopolynomial time O($nT$). This time has been very significantly improved but remains pseudopolynomial with functions of $n$ giving O($f(n)T$) [KX16], [BR17], [JVW20]. The use of modular arithmetic in the context of the subset sum problem is quite widely described, for example in [C23]. The selection of one algorithm from a set of algorithms to solve the related partition problem, selected using initial inputs, is described by Korf et al[KSM14]. 

\section {The selective algorithm system}

The system described solves this form of the subset sum problem. It includes several standard elements of dynamic programming. These elements include an array $s_x$ ($x$:$1..T$) where $s_x$ = index of the value that first computed the subset sum $x$, the rejection of candidate subsets that repeat a previously computed sum and the recovery of the solution subset by backtracking from $s_T$. One new element is the mapping of sums into bins. The bin data references sums classed as either computed ($s_x >0$) or uncomputed ($s_y=0$). 

The core issue tackled by this system is avoiding the computation of subset sums which duplicate already computed sums. A subset sum algorithm's efficiency $e = min(T/z, 1)$, where $z$ is the number of sums computed, including duplicate sums. $e=1$ corresponds to no duplicates being computed. $e=1/n$ corresponds to $n$ duplicates being computed for each sum.  O($T$) algorithms have $e \approx 1$ while  O($nT$) algorithms have $e \approx 1/n$.  Mapping subset sum space into bins makes it easier to avoid duplicates. 

The classic dynamic programming algorithm (DP) applied to subset sum adds $v_i$ to each known computed sum $x$ to see if the candidate sum $y=x+v_i$ is uncomputed by checking if $s_y=0$. This can be done in reverse. The subtractive dynamic programming algorithm (SDP) subtracts $v_i$ from a known uncomputed sum $y$ to check the sum $x=y-v_i$. If $s_x> 0$ then the sum $x$ has been previously computed and $y$ is the sum for a newly computed subset. This newly computed subset has been formed by adding $i$ to the previously computed subset whose sum = $x$.

Let $e*$ be the bin level efficiency of an algorithm applied to a value $v_i$. $e*=a/b$ where $a$ is the number of different sums created for this bin and value and $b$ is the number of sums considered for this bin and value. When local sum space is empty DP has $e*=1$ and SDP has $e*=0$.  When local sum space becomes half full, DP and SDP both have $e* \approx 0.5$  When local sum space is full, DP has $e*=0$ while SDP has $e*=1$. For each $v_i$ and each bin, this system selects the more efficient of DP or SDP. Assuming random distributions of sums in bins, $e > 0.5$ for this system. This corresponds to computational O($T$) for value processing.

This computational O($T$) must be combined with the computational complexity of bin initialisation and management. With $kn$ bins and $k$ small ($k=2$ was chosen), bin management has $n$ values each checking $kn$ bins, giving O($n^2$). Bin initialisation has computational O(max($T$, $n$)) since $T$ entries in a fixed array are created in a loop $1..T$ along with initialising $kn$ bins.  Sorting the input values has the lower complexity O($nlogn$). Accordingly the system has computational O(max($T$, $n^2$)), assuming random distributions of sums in bins. The $kn$ bins and two associated $T$ arrays have total space O(max($T$,$n$)).

\subsection {Process}

The system initially sorts $v_i$: $i=1..n$ low to high. All entries in a fixed array linked list of uncomputed sums $1..T$ are now filled, each entry linked to its two neighbours.  The $g$ bins are then initialised from this list of uncomputed sums ($g=kn$). The bin data is described below (3.1.1).

After bin initialisation, this system processes each value $v_i$ in turn. During this value processing a series of candidate subset sums $y$ are generated. The system takes the actions described below (3.1.2) for each candidate sum.
For each $v_i$, if $v_{i-1} = v_i$, repeated value dynamic programming (RVDP: 3.1.6) is used. Otherwise, for each bin $j=1..g$ the probably most efficient of DP or SDP is selected (3.1.7) and applied.
At the end of processing each $v_i$ the system takes the appropriate actions (3.1.2) on the candidate sum $y=v_i$.

\subsubsection {Bin data}

Each bin up to the last handles the same number of sums $l$ where $l=T/g$. (A minimum of $l=500$ was used in testing.)
Each bin $j$'s data describes the subset sum distribution over the range of sums $(j-1)l+1$ .. $jl$. The exception is the last bin $T/l$ which has the upper range limit $T$.
Each bin has the following data.

1 lowest and highest computed sum in its range (needed for SDP).

2 appended to array of computed sums in its range (needed for DP).

3 count of computed sums in its array and so over its range (needed in DP and to choose between DP and SDP).

4 lowest and highest uncomputed sum in its range (needed for SDP and to choose between DP and SDP).

\subsubsection {Actions taken on each candidate sum}

The candidate sum $y$ will be defined as newly computed if $s_y=0$. 

1 if $s_y=0$, take the actions below (3.1.3) for this new computed sum.

2 if $s_y> 0$, this is a duplicate sum and ignored.

\subsubsection {Actions taken on each new computed sum}

 Actions taken on the sum $x$ newly computed while processing $v_i$.

1 $s_x \gets i$. 

2 The data in bin $d=x/l$ is updated. $x$ is appended to the array of computed sums in bin $d$, the count of computed sums in $d$ is incremented and the computed and uncomputed sum limits in bin $d$ may be changed.

3 $x$ is appended to $h$ ($h$ is the list of new subset sums created at $v_i$). $h$ may be used by RVDP for $v_{i+1}$ (3.1.6).

4 $x$ is removed from the uncomputed sums' linked list. This linked list is used in SDP (3.1.5).

\subsubsection {DP: dynamic programming}

DP processes the entries in the array of computed sums in bin $j$. The length of this array is the number of computed sums in bin $j$.

1 DP considers each computed sum $x$ in this array to generate candidate sums $y=x+v_i$.

\subsubsection {SDP: subtractive dynamic programming}

SDP processes the range of uncomputed sums corresponding to the range of computed sums stored in bin $j$. It considers each uncomputed sum $y$ in turn. The uncomputed sum $y$ will be defined as a new computed sum
 if $s_{y-v_i}> 0$. This range of uncomputed sums is calculated and processed as follows.

1 Let  $a1=($lowest computed sum in bin$_j)+v_i$.

2 Let $a2=($highest computed sum in bin$_j)+v_i$.

2 Bins $b1$ and $b2$ contain the sum data for the range of uncomputed sums $a1..a2$: $b1=a1/l$ and $b2=a2/l$.

3 SDP starts from the highest uncomputed sum in bin $b1$ and considers all uncomputed sums down to $a1$, stepping down the uncomputed sums' linked list. 

4 SDP then starts from the lowest uncomputed sum in bin $b2$ and considers all uncomputed sums up to $a2$, stepping up the uncomputed sums' linked list.

\subsubsection {RVDP: repeated value dynamic programming}

RVDP only needs to process the subset sums in the list $h$ created at $v_{i-1}$, as all other candidate sums for this value were already considered at $v_{i-1}$. 

1 RVDP adds $v_i$ to each computed subset sum $x$ in $h$ to generate candidate sums $y=x+v_i$.

\subsubsection {Choosing the probably most efficient of DP or SDP}

For this $v_i$ and this bin $j$, the system estimates the number of subsets to be computed by DP ($c1$) and by SDP ($c2$). It chooses DP if $c1 < c2$ and SDP otherwise. $c1$ is the number of computed sums in bin $j$ (3.1.4).
$c2$ is estimated as follows.

As previously defined in 3.1.5, $a1..a2$ is the range of candidate uncomputed sums corresponding to the range of computed sums stored in bin $j$. $b1$ and $b2$ are the respective bins for the sums $a1$ and $a2$.
If used, SDP would work down from the highest remaining uncomputed sum in bin $b1$ to sum $a1$ and up from the lowest remaining uncomputed sum in bin $b2$ to sum $a2$.

So, $c2 \approx c3 + c4$ where

$c3$ is the maximum number of uncomputed sums which need to be checked in bin $b1$. So, $c3= $(highest remaining uncomputed sum in bin $b1$)$ - a1$.

$c4$ is the maximum number of uncomputed sums which need to be checked in bin $b2$. So, $c4= a2 - ($lowest remaining uncomputed sum in bin $b2$).

\subsubsection {Heuristic}

The folowing heuristic reduces runtime significantly for some cases. It has no effect on efficiency.
Not all the $g$ bins need be processed for each $v_i$ so the system processes a list of useful source bins. Useful source bins have at least one computed sum and have at least one candidate bin with at least one uncomputed sum. A list of such useful source bins is recreated as appropriate, depending on the current $v_i$ and the value when this list was last created. The list is created either directly from a list of bins which have at least one computed sum or indirectly from a linked list of bins which still have at least one uncomputed sum, whichever way will be the fastest. Bins which now have one computed sum are appended to this useful source bin list after its creation.

\section {Special handling of some input value cases}

Two classes of input value cases generate subset distributions on which this system would become inefficient. Such distributions have bins which never become full and generate subset sums with a high percentage of duplicates. One class of these input value cases is characterised by having many values that share a common divisor (eg: almost all even values). Cases in this class are recognised and handled by iterative processing of significant common divisors (IPSCD), described below (4.1).

The second class of input cases is recognised by an unacceptably low efficiency during value processing. This class is handled by partitioning the values into those already processed and those still to be processed. This class is chararcterised and recognised by two features. The first feature is $e < P1$, where $e$ is the efficiency so far and $P1$ is a constant. Define  $e*_i=a/b$ where $a$ is the number of different subset sums created and $b$ is the number of subset sums considered for $v_i$. The second feature is that the average $e*_i < P2$, where this average is measured over the last $P3$ values and $P2$ and $P3$ are constants.  ($P1=0.2$, $P2=0.01$ and $P3=50$ in testing).  If this class is recognised, the system breaks $v$ into two partitions, similar to the partition action in Horowitz and Sahni[HS74]. Partition $A$ contains the values already processed and partition $B$ contains the values yet to be processed. Array $s$ is copied to an array $sA$ and the system reinitialised. Partition $B$ values are now processed, defining success when either $x=T$ or $sA_{T-x}> 0$, for each new subset sum $x$. Otherwise, partition $B$ is processed as if it was an initial input case, except it cannot be further partitioned. Partitioning has cost O(max($T$, $kn$)) from copying and clearing the $s$ array and reinitialising the $kn$ bins.

\subsection {Significant common divisors}

IPSCD recognises and efficiently processes input cases whose values have significant common divisors: for example, a case with all even values. In this example IPSCD determines that all values share the divisor 2. From this, it determines which sums are computable (the even ones) and generates the uncomputed subset sum linked list ignoring the uncomputable sums (the odd ones). This means the gaps in uncomputable sums are ignored and the efficiency maintained. The algorithm used for recognising the significant greatest common divisor is given below (4.1.1).

A case can also have multiple partially overlapping sets of values, each set categorised by all its values sharing the same common divisor.  IPSCD recognises each set's common divisor in turn. The first set's values are processed independantly of the rest of the values. Sums which are uncomputable from this divisor are ignored during this processing. Then the IPSCD algorithm detects the second shared divisor and this set's values are processed independantly of the rest of the remaining values, merged with the subset sum array data from the first set. Sums which are uncomputable from this divisor and which have not already been computed are ignored during this processing. IPSCD continues to iterate until it recognises no significant shared divisor in the remaining values and processes the remaining values ignoring no sums. 

IPSCD steps from $1..T$ by its current divisor to generate the uncomputed subset sum linked list. Therefore the extra computational cost of processing multiple divisors is less than O($T(...1/8 + 1/4 + 1/2)$), which is O($T$) complexity.

\subsubsection {Recognising the significant greatest common divisor}

The lowest few unprocessed different values are chosen for this analysis (40 in testing). Let GCD be the standard greatest common divisor algorithm. 

1: find all greatest common divisors $d$ between all pairs of chosen values. If processing the residual values after applying a previous divisor $x$: each $d= GCD(d,x)$. 

2: $c_d$ equals the number of chosen values for which $(v_i$ mod $d) = 0$.

3: reject all divisors with $c_d >$ limit to ensure some level of significance of this divisor for these chosen values. This limit was set at 20 in testing for all divisors except $d=2$, whose limit was set at 32. 

4: select the divisor $d$ with the highest probabilistically adjusted count: $dc_d$.

\section {Experimental validation}

For validation, this system was implemented in 1000 lines of C++ plus 1500 lines for testing. 
This was run on an 8Gb 4-core 3.5ghz Windows 10 machine.  Only one core was used in the tests. There was no parallel logic and no GPU use. The system was extensively cross checked against a variant of the classic algorithm of Horowitz and Sahni [HS74] for $n <= 45$.  

\subsection {Results}

   The system was tested for $n=10^2..10^7$ and $T=10^5..10^8$ on cases with randomly generated values in the range $1..T/4$. 10 trials were performed on each $n,T$ combination. The tests were done not stopping on success, to more clearly show the system's experimental computational load. The two tables report data from the same trials.

   Table 1 shows the average efficiency, with respect to $n$ and $T$. The efficiency is approximately independant of $n$ and $T$. Computational O($T$) algorithms have $e \approx 1$ while O($nT$) algorithms have $e \approx 1/n$. This efficency result is compatible with O(max($T$, $n^2$)) and incompatible with O($f(n)T$) for any significant $f(n$).  

   Table 2 shows the average runtime, excluding input value creation and sorting, for each $n,T$ combination. This table shows the runtime is dominated by $T$, not by $n$. 

  The fraction of all operations in the system which were bin operations was never more than 0.70 in any test.

\subsubsection {Possible worst cases}

   One hundred different candidate worst case value multisets were constructed and experimentally checked. These cases had various mixtures of subsets of values with different features designed to stress the system. These features included generation of dense or sparse regions of subset sums, multiple subsets of values with different shared greatest common divisors, varying gaps in subset sum regions and repeated patterns of values. The system was tested on all these candidate worst cases for $n=10^2..10^7$ with $T=10^8$. The worst case time was 207.9 seconds and worst case efficiency was 0.073, both recorded at $n=10^7$. These worst case results are about 5 times slower and 12.5 times less efficient than the average for random value cases with the same $n=10^7$ and $T=10^8$. Should the system's processing of any of these candidate worst cases approach computational O($nT$), these expected worse case results would have been of order $10^7$ times slower and $10^7$ times less efficient.

\subsection {Discussion and conclusions}

The experimental results show that bins present information useful to the subset sum problem. The use of a fixed array linked list of uncomputed sums and the use of bins as a form of index to this linked list is not worse than computationally O(max($T$, $n^2$).
Choosing the number of bins to be $2n$ is effective. However, other choices are possible, such as $n^k$ for low $k$. 

 Three other bin design choices were considered and the first one of these coded and tested.  The first  was to merge neighbouring bins with no remaining uncomputed sums.  However, some of the design details for this are very complex for almost no tested improvement. The second choice was that bins could have variable sizes. Variable size bins probably need secondary support, such as a sum to bin mapping array but may be easier to mathematically analyse. The third choice was to have fixed size bins, independant of $n$. However, if $l$ is the chosen fixed bin size, this generates $T/l$ bins and so tends to produce computational complexity O(max($T$, $nT/l$)).

Subtractive and repeated value dynamic processing are validated as more efficient than standard dynamic processing under some conditions.

Computational efficiency measured over all the input values can be optimised by selecting algorithms by their forecast local efficiency, the selection being done per bin per value. The estimation of DP and SDP local efficiency from the exact or estimated number of subsets they would process is acceptable.

The experimental results validate this system's computational complexity O(max($T$, $n^2$)) but this result remains unproven. For example, the assumption of random distributions of sums in bins appears acceptable but remains theoretically unsupported. Again,
the use of dynamic partitioning to reduce some cases to acceptable efficiency relies on all causes of unacceptable inefficiency being in the first partition's values. No candidate worst case was found that failed to be handled by this partitioning, but this assumption also remains theoretically unsupported.

\begin{table}[H] 
\caption{Average efficiency over 10 trials}
\centering
\begin{tabular}{lcccccr}
\hline\hline\\
$T$& $n=10^2$  & $n=10^3$  & $n=10^4$ & $n=10^5$  & $n=10^6$ & $n=10^7$\\
\hline
$10^5$ & 0.86 & 0.80 & 0.75 & 0.74 & 0.76 & 0.79 \\
$10^6$ & 0.91 & 0.80 & 0.90 & 0.83 & 0.77 & 0.79 \\
$10^7$ & 0.92 & 0.77 & 0.89 & 0.96 & 0.89 & 0.78 \\
$10^8$ & 0.94 & 0.77 & 0.83 & 0.96 & 0.98 & 0.91 \\
\hline 
\end{tabular}
\label {table:nonlin}
\end{table}

\begin{table}[H]
\caption{Average runtime over 10 trials excluding sort time (secs)}
\centering
\begin{tabular}{lcccccr}
\hline\hline\\
$T$& $n=10^2$  & $n=10^3$  & $n=10^4$ & $n=10^5$  & $n=10^6$ & $n=10^7$\\
\hline
$10^5$ & 0.025 & 0.025 & 0.044 & 0.082 & 0.103 & 0.308 \\
$10^6$ & 0.235 & 0.242 & 0.259 & 0.435 & 0.849 & 1.052 \\
$10^7$ & 2.243 & 2.158 & 2.238 & 2.398 & 4.003 & 7.596 \\
$10^8$ & 25.38 & 21.31 & 21.25 & 22.39 & 25.05 & 42.84 \\
\hline
\end{tabular}
\end{table}

\end{document}